# Understanding Shannon's Entropy metric for Information

Sriram Vajapeyam
v.sriram.blr@gmail.com

24 March 2014

## 1. Overview

Shannon's metric of **"Entropy"** of information is a foundational concept of ***information theory*** [1, 2]. Here is an intuitive way of understanding, remembering, and/or reconstructing Shannon's Entropy metric for information.

Conceptually, ***information*** can be thought of as being stored in or transmitted as *variables* that can take on different *values*. A *variable* can be thought of as a unit of storage that can take on, at different times, one of several different specified values, following some *process* for taking on those values. Informally, we get information from a variable by looking at its value, just as we get information from an email by reading its contents. In the case of the variable, the information is about the process behind the variable.

The ***entropy*** of a variable is the **"amount of information"** contained in the variable. This amount is determined *not just by* the number of different values the variable can take on, just as the information in an email is quantified not just by the number of words in the email or the different possible words in the language of the email. Informally, the amount of information in an email is proportional to the amount of "*surprise*" its reading causes. For example, if an email is simply a repeat of an earlier email, then it is not informative at all. On the other hand, if say the email reveals the outcome of a cliff-hanger election, then it is highly informative. Similarly, the information in a variable is tied to the amount of surprise that value of the variable causes when revealed.

**Shannon's entropy** quantifies the amount of information in a variable, thus providing the foundation for a theory around the notion of information.

**Storage and transmission of information** can intuitively be expected to be tied to the amount of information involved. For example, information may be about the outcome of a coin toss. This information can be stored in a Boolean variable that can take on the values 0 or 1. We can use the variable to represent the **raw data** corresponding to the coin toss, viz., whether the coin toss came up heads or not. In digital storage and transmission technology, this Boolean variable can be represented in a single "bit", the basic unit of digital information storage/transmission. However, this bit directly stores the value of the variable, i.e. the **raw data** corresponding to the outcome of the coin toss. It *does not succinctly capture* the information in the coin toss, e.g., whether the coin is biased or unbiased, and, if biased, how biased.

Whereas, **Shannon's entropy** metric quantifies, among other things, the *absolute minimum* amount of storage and transmission needed for succinctly capturing any **information** (as opposed to raw data), and in typical cases that amount is *less than* what is required to store or transmit the raw data behind the information. Shannon's Entropy metric also suggests a way of representing the information in the calculated fewer number of bits.

The figure below gives a conceptual overview of this article.

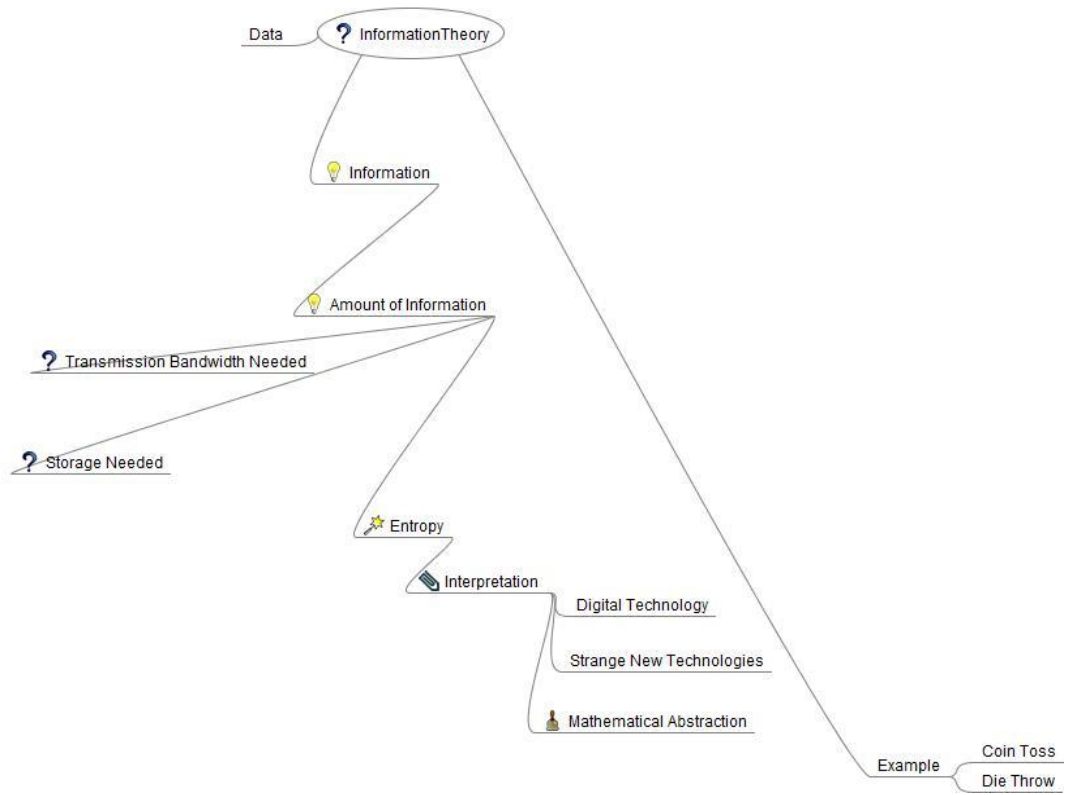

## 2. Meaning of Entropy

At a conceptual level, Shannon's Entropy is simply the **"amount of information"** in a variable. More mundanely, that translates to the **amount of storage** (e.g. number of bits) required to store the variable, which can intuitively be understood to correspond to the amount of information in that variable. The calculation of this number and therefore the amount of information in a variable is more involved than might appear at first sight; specifically, it is not simply the number of bits required to represent all the different values a variable might take on, which is just the raw data.

For example, a variable may take on any of 4 different values. In digital storage, 2 bits would be sufficient to uniquely represent the 4 different values, and thus the variable can be stored in 2 bits. However, this is an upper limit on the required storage; it is the amount of storage required to store the raw "**data**" of the variable, not the "**information**" in that data. Less storage *might be* sufficient to store the information, depending on the process by which the variable takes on different values. Shannon's entropy metric helps identify that amount of storage needed for the information. One alternative way of looking at entropy is thus as a measure of "compressibility" of the data, i.e., a **compression metric**: how much can the raw data of a variable be compressed without losing the information in the variable?

### 3. Concept of "Amount" of Information

Intuitively, one way to understand the concept of "amount of information" in a variable is to tie it to how **difficult** or **easy** it is to **guess that information** without having to look at the variable: the easier it is to guess the value of the variable, the less "surprise" in the variable and so the less information the variable has. Another way to view information is to contrast it with the amount of data. For example, two different Boolean variables could be stored in 1 bit each, but the amount of information in the two may be quite different, as illustrated below.

Suppose a coin is completely *biased* and always comes up heads when tossed, then the *random variable* representing the coin toss's outcome has *probability* 1 of coming up heads (in other words, it is a constant), and thus there is no need to store or transmit that variable as it can be trivially guessed at any time. Thus, the amount of information in that variable is zero.

On the other hand, if we had a perfect coin with half-half chances of coming up heads or tails upon a coin toss, then we can guess the outcome of a toss with only 50% accuracy (probability 0.5), so it is necessary to store/transmit the actual value of that coin toss outcome's random variable in order to know its value with better than 50% accuracy. Thus, the amount of information in this second random variable is much higher than in the first case.

Contrast the fact that, for both of the above coins, the raw data regarding their toss outcomes need 1 bit each to be stored.

Instead of a coin, suppose we had a perfect die (cube) with 6 possible outcomes on the toss (roll) of the die. The amount of information in the corresponding random variable is even higher, as it now even harder to guess the outcome of the die roll, we have only 1/6th chance of guessing the outcome correctly -- much less than 50%.

### 4. Quantifying the Amount of Information

Now, how do we quantify the "amount of information" in a random variable? One way to represent this "amount of information" is the number of bits it takes to represent/express the variable. In a naive representation of variables, if a variable can take on only two values, it can be represented with just 1 bit; if it can take on any of 4 values, 2 bits are needed (the 2 bits can be said to form a "word"); if 9 different values, then 4 bits are needed; etc. But this is the storage required for the raw data, not for the information content of the data.

In a more sophisticated representation of the variable, if a variable is easier to guess, then we can leverage that fact to reduce the number of bits needed to store/transmit that information. For example, if a die is 80% likely to come up with the number "3", then one way to leverage the

"guess-ability" of the die is to store/transmit only a single bit with the value 0 whenever the die actually comes up with "3", and store/transmit more bits, starting with a first bit of value 1, when the die comes up with other values. The single bit transmission (of bit value "0") simply indicates to the receiver that the dice was tossed; the receiver then just maps that information to the information that the toss outcome was "3", without needing the transmitter to actually specify that value! Further, when the first bit is a 0, the receiver knows not to look for more bits associated with this particular storage/transmission. This avoids the need for an "end" marker for each stored or transmitted "word", which would be additional cost otherwise.

All this is an intelligent form of "coding" (or, "encoding") the information; it leverages the probabilities or bias to reduce the average size of the code. While in a naive representation we would need 3 bits to represent each outcome of a die having 6 faces, by leveraging probabilities we use only 1 bit whenever the die comes up with "3" -- a more frequently occurring event -- thus reducing the average number of bits needed (over multiple dice throws) to store the die throw outcome. This reduction happens even if we are forced to use more than 3 bits for some infrequently-occurring outcome of the die throw.

One can thus leverage the probabilities of the different values to reduce the number of bits needed -- if and only if the variable has a non-uniform distribution. If not, there will be no reduction in the number of bits needed. The intuition here is that if a variable is more likely to take on one value than another, then it is easier to guess the value of the variable without looking at it, thus the variable has less information in it, and thus it takes fewer bits to store/transmit the value.

### 5. Example of Information Quantity

Let us look at a more detailed but simple, concrete example.

Suppose a variable can take on 3 different values a,b,c, but half the time it takes on the value 'a', and a quarter of the time each the values 'b' and 'c'. Now, we can represent 'a' with just one bit having the value "0"; 'b' with 2 bits "10", and 'c' with 2 bits "11". This is Huffman Coding. Observe that decoding the representation is easy, and it does not need "end-of-representation" markers. Specifically, when the first bit is a 0, the receiver knows to stop reading that "word" right there; when the first bit is a 1, the readers knows to also read the next bit to complete the "word".

With the above representation, what is the number of bits needed to represent this variable?! Well, we need just 1 bit half the time (when the value taken on is 'a'), and 2 bits each the other 2 times ((when the value taken on is either 'b' or 'c'), so the average number of bits needed is {(0.5*1) + (0.25*2) + (0.25*2)} = 1.5bits!

So, the entropy of the above variable having those specified probabilities of taking on different values is 1.5!

**6. The Entropy Formula**

Now, to understand the entropy formula, let us write down the three probabilities in the above example (section 5) for the occurrences of a, b, and c as follows:
- p(a) = 0.5  = 2/4
- p(b) = 0.25 = 1/4
- p(c) = 0.25 = 1/4

The key thing to notice here is that we have written p(a) as 2/4 and not 1/2, thus ensuring that all the three probabilities have the same denominator. Writing it this way leads us to an interpretation of the Entropy formula.

When p(x==c)=1/4 (i.e., p(c)), the numerator 1 and denominator 4 say that the value 'c' is taken on once in 4 times, which means there might be up to 3 other values that the variable can take on at other times (i.e., when it does not take on the value c). In which case, there are 4 values altogether and we will need the following number of bits to represent them all:
- $\log_2(4) = \log_2(1/p)$ bits = 2 bits

So, 'c' will need 2 bits to be represented as distinct from those other values.

To represent the one 'b', we will similarly need $\log_2(4) = 2$ bits.

For 'a', we can interpret p(a) = 2/4 as the value 'a' being taken on twice in four times. Or, we can now say that of the 4 values the variable can take on, two are 'a's, one is 'b', and one is 'c'. Now, to represent the two 'a's among the four values, we don't need two different storage representations, as the two 'a's are the same. So, from the overall number of bits required to represent the two 'a's of the variable, we can deduct $\log_2(2)$ bits, since presumably those $\log_2(2)$ bits were meant to differentiate the two 'a's which are actually identical. Thus, to represent 'a', we will need:
- $\log_2(4) - \log_2(2) = 1$ bit

Now, to determine the overall amount of storage required for the variable, we simply add up the above storage requirements (for the different individual values of the variable) in proportion to their frequencies of occurrence:
- 'a' occurs half the time and needs 1 bit: **0.5*1**
- 'b' occurs a quarter of the time and needs 2 bits: **0.25*2**
- 'c' occurs a quarter of the time and needs 2 bits: **0.25*2**
- In total, we need **0.5*1 + 0.25*2 + 0.25*2** = 1.5 bits

But, this is exactly Shannon's formula [1] for Entropy!
- ***Shannon Entropy* E = $-\sum_i (p(i) \times \log_2(p(i)))$**

Note that the minus sign takes care of the fact that p(i) is a fraction. For example, for 'a',
- $-p(a) \times \log_2(p(a)) = -\{0.5 \times \log_2(2/4)\} = -\{0.5 \times [\log(2)-\log(4)]\} = +\{0.5 \times [\log(4)-\log(2)]\} = 0.5 \times 1$ **!!**

Thus, entropy is a direct measure of the number of bits needed to store the *information* in a variable, as opposed to its raw *data*. Thus, entropy is a *direct measure of the "amount of information"* in a variable.

### 7. Mathematical Approximation of Required Storage

So far so good, but the intuition break down a bit when we have probabilities that have numerators or denominators that are not powers of 2. For example, suppose 'a' occurs with probability 9/27, that means there might be 27 different values the variable can take on, and 'a' is 9 of them. Now, $\log_2(27) = 4.75488..$, and that is not an *integral* number of bits! So how do we interpret $\log_2(p)$ in this scenario? Obviously we can't say that we need $\{\log_2(27)-\log_2(9)\} = (4.75488-3.1699) = 1.5849$-odd bits to store 'a'! In digital storage, bits come whole, in counts of integers, and not in fractions!

In this case, the entropy formula becomes a **mathematical entity**, perhaps having no exact real-world analogue in today's world. However, to understand the results of the Shannon's Entropy formula in these scenarios, we can imagine strange new kinds of technology where storage units can come in non-integral amounts and can take on non-integral values. Such technologies may not be all that strange after all: for example, in quantum computers, a "**qubit**" is a single bit that can *simultaneously* have the values 0 and 1 with different probabilities α and β (or something like that, anyway!). A little more mundanely, in the above example, suppose we had tri-state logic where each storage unit could take on one of 3 possible values. Now, if we change the base of the logarithm in the entropy formula to 3, we suddenly have $-p*\log_3(p) = -(9/27)*\log(9/27) = (9/27)*\{\log(27)-\log(9)\} = (9/27)*\{3-2\} = 1/3$. That is, 'a' can be represented in one unit of tri-state storage, and 'a' occurs 1/3-rd of the time!

Thus, a simple change in the base of the logarithm used in the entropy formula tells us the absolute minimum number of bits required to store the information in a technology where each storage unit allows as many states as the base of the logarithm. In any case, even for digital storage having binary bits, Shannon's Entropy represents a **lower-bound** on the number of actual bits required to store or transmit information. In the above example, we will need 2 bits to represent 'a', 2 being the next integer higher than 1.263.